IAC-11-B3.3.6

# THE UNITED NATIONS HUMAN SPACE TECHNOLOGY INITIATIVE (HSTI)

**Mika Ochiai**
Office for Outer Space Affairs, United Nations Office at Vienna, Austria, mika.ochiai@unoosa.org

**Ingrid Dietlein, Aimin Niu, Hans Haubold, Werner Balogh, Takao Doi**
Office for Outer Space Affairs, United Nations Office at Vienna , Austria
ingrid.dietlein@unoosa.org  aimin.niu@unoosa.org  hans.haubold@unoosa.org
werner.balogh@unoosa.org  takao.doi@unoosa.org

The Human Space Technology Initiative (HSTI) has been launched under the framework of the United Nations Programme on Space Applications with its aims of promoting international cooperation in human spaceflight and space exploration-related activities; creating awareness among United Nations Member States on the benefits of utilizing human space technology and its applications; and building capacity in microgravity education and research. The International Space Station (ISS), being operational with a permanent crew of six (6), is an unprecedented facility for research on science and technology and can be regarded as one of the greatest resources for humankind to explore space. The HSTI seeks to promote human space technology and to expand ISS utilization. This report describes the background, objectives, and current three-year work plan of HSTI which is composed of organizing expert meetings, seminars, and workshops as well as building capacity by distributing educational materials and zero-gravity experiment instruments.

## I. INTRODUCTION

In 1959, the United Nations General Assembly created the Committee on the Peaceful Uses of Outer Space (COPUOS) in order to review the scope of international cooperation in peaceful uses of outer space, to devise programmes in this field to be undertaken under the United Nations' auspices, to encourage continued research and dissemination of information on outer space matters, and to study legal problems arising from the exploration of outer space.

At the first United Nations Conference on the Exploration and Peaceful Uses of Outer Space (UNISPACE '68) [1], held in Vienna in 1968, when Member States recommended the creation of a dedicated programme in the framework of the United Nations to support countries which lacked the human, financial and technical resources necessary to fully utilize the benefits of space technology. In its resolution 2601 A (XXIV) of 16 December 1969, the General Assembly endorsed "the recommendation of the Committee on the Peaceful Uses of Outer Space for the appointment by the Secretary-General of a qualified individual with the full-time task of promoting the practical applications of space technology" [2]. Initial activities - in what was later to be called the "United Nations Programme on Space Applications" - commenced in May 1971. The United Nations Office for Outer Space Affairs (UNOOSA), as the Secretariat for COPUOS, was also given the responsibility for implementing the Programme.

Following the second UNISPACE conference held in 1982 (UNISPACE '82), the mandate of the Programme was broadened [3]. The third and the latest UNISPACE Conference, held in 1999 (UNISPACE III), touched upon, for the first time, the significance of the International Space Station (ISS) for humanity and the importance of encouraging international cooperation, particularly cooperation and partnerships between countries operating and already using the ISS and developing countries [4].

The ISS has been the only permanently inhabited human outpost in space until now for about a decade. It is 110 meters long, 74 meters wide, weighs almost 420 metric tons, composed of modules and parts from the United States of America, the Russian Federation, Canada, Japan and Europe. It combines the competences of these nations to provide an extraordinary facility for the purpose of enhancing knowledge through science and technology performed in space. It is an outstanding example of international cooperation on a large scale for a single and shared purpose.

The Human Space Technology Initiative (HSTI), launched by UNOOSA in 2010 in the framework of the United Nations Programme on Space Applications, seeks to promote active participation of countries in research and education in microgravity science and technology and to expand utilization of the ISS by non-ISS partners.

A work plan spanning three years has been developed in order to meet the HSTI mandates and objectives. The current and future activities of HSTI are presented and discussed in detail.

## II. THE INTERNATIONAL SPACE STATION (ISS) AND ITS UTILIZATION

### II.I Overview of the ISS [5, 6]

In 1998, the construction of the International Space Station (ISS) was begun by a partnership of five space agencies, the Canadian Space Agency (CSA), the European Space Agency (ESA), the Ministry of Education, Culture, Sports, Science and Technology (MEXT) of Japan along with the Japan Aerospace Exploration Agency (JAXA), the National Aeronautics and Space Administration (NASA) of the United States of America, and the Russian Federal Space Agency (Roscosmos), representing 15 countries. It was the first ever multipurpose manned space facility in low Earth orbit. Since 2009, the ISS has been operational with a permanent crew of six (6) and now the ISS assembly is almost complete.

The major intended purpose of the ISS is to provide an Earth-orbiting research facility that houses experiment payloads, distributes resource utilities, and supports permanent human habitation to conduct science and research experiments in a microgravity environment.

The U.S. laboratory, also known as Destiny, is the major U.S. contribution of scientific capacity to the ISS. It provides equipment for research and technology development and houses all the necessary systems to support a controlled environment laboratory. Destiny provides a year-round, shirtsleeve atmosphere for research in areas such as life science, physical science, material science, Earth and space science.

Meaning "dawn" in Russian, the Rassvet Mini-Research Module-1 is used for science research and cargo storage as well as for providing an additional docking port for the Russian Soyuz and Progress vehicles at the ISS. This facility serves as a home to biotechnology and biological science experiments, fluid physics experiments, and education research. The Poisk Mini-Research Module-2 is a multipurpose extension of the ISS, which provides an additional docking port for visiting Russian spacecraft and serves as an extra airlock for spacewalkers wearing Russian Orlan spacesuits.

ESA has contributed the Columbus module - a research laboratory which supports sophisticated multi-disciplinary research, having internal and external accommodation for dozens of experiments in life and physical sciences, space and Earth science, technology, commercial R&D, education, and finally, human exploration preparation.

The Japanese Experiment Module Kibo, which is Japanese for "hope," is Japan's first human space facility and its primary contribution to the ISS. The Kibo laboratory consists of a pressurized module and an exposed facility which have a combined focus on space medicine, biology, biotechnology, astronomy, Earth and space science, material production, and communications research.

The completed ISS will put its full capabilities to work as an unprecedented facility for multidisciplinary research and technology development, and an outpost to conduct human exploration. As a unique symbol of international collaboration, the ISS can be regarded as one of the greatest resources or assets for humankind to improve life on Earth and to explore space.

### II.II The ISS-Enabled Educational Activities

Space technology and particularly human space technology can produce a positive effect also on the education of young engineers and scientists for preparation for the future. In addition, capacity building and training of local scientists and researchers is the prerequisite for realizing international cooperation in microgravity science. Being the largest complex ever built in space, the ISS has also helped students become more excited about space and get them interested in science and technology.

The ISS partners have conducted respective initiatives on education concerning space research and increasing education and outreach in general at all levels. Those include organizing events offering learning opportunities about the Earth, amateur radio on the ISS, audio/video interaction between the ISS crew and students, demonstrations of the behaviour of simple items in microgravity, and conducting a number of zero-gravity experiments selected by the general public.

CSA offers mathematics, science, and technology learning products to students and educators by providing direct access to CSA scientists, engineers, and astronauts in order to increase educator awareness of the space context as well as tools and scenarios. The CSA Education and Outreach learning Programmes provide unique learning experiences that encourage hands-on space-focused science learning.

ESA offers primary and secondary schools online lessons about various topics, including the Columbus Module, ISS education kits containing fully illustrated information, sources, and experiments conducted on the ISS, and radio contact between astronauts on the ISS and students (ARISS). The SUCCESS program (Space Station Utilisation Contest Calls for European Student initiativeS) allows university-level students to submit ideas for payloads which are reviewed by scientists at ESA for scientific merit and feasibility.

JAXA has conducted experiments on the mutation effects of spaceflight on seeds and studied the ecosystem and the Earth's environment by growing spaceflight plant seeds through JAXA's Spaceflight Kids Space Mission I. The seeds were then distributed to the schools. In order to increase public awareness, JAXA periodically conducts a number of zero-gravity experiments selected by the general public.

NASA has organized events that have involved over 31 million school students around the world, offering learning opportunities concerning Earth knowledge, Amateur Radio on the ISS (ARISS), audio/video interaction between the ISS crew and the students, and demonstrations of the behaviour of simple items in microgravity. Students have conducted classroom controls of ISS experiments and investigated autonomous rendezvous technologies, and students have competed to design experiments to be conducted on the ISS.

Roscosmos conducts experiments and lessons from space to popularize space research, to promote achievements of cosmonautics, and to involve students in software/hardware development and other experiments in an amateur radio network and physics.

In addition to these activities conducted by the ISS partners, in order to share best practices and unite efforts to foster interest in space, science, and technology among students worldwide, the Centre National d'Etudes Spatiales (CNES) of France, CSA, ESA, JAXA, NASA, and non-agency partners involved in space-related educational pursuits created the International Space Education Board (ISEB).

II.III Utilization of the ISS - Coordination of Research Issues and Non-Partner Participation Policy

In order to optimize cooperation and avoid duplication in the utilization of the ISS laboratories, international working groups for coordination of ISS utilization were created within the ISS partnership. The working groups are assisted by the ISS User Operations Panel (UOP) for the coordination of strategic planning and program research objectives, and by the ISS Program Science Forum (PSF) for the coordination of research issues, research communication, and results tracking.

One possible framework for utilizing the ISS by non-ISS partner countries is offered by the Non-Partner Participation Policy (NPP). In 2002, the ISS partnership developed this Non-Partner Participation Policy (NPP) which governs how non-ISS partners can participate in the ISS programme. After an initial bilateral agreement between a non-ISS partner participant with one of the five partners of the ISS, the ISS partnership will review the proposed bilateral cooperation for approval. Non-ISS partners are encouraged to cooperate with ISS partners to discuss research proposals. Through the year 2010, a total of 58 countries have participated in ISS utilization [7].

III. HUMAN SPACE TECHNOLOGY INITIATIVE (HSTI)

III.I The United Nations Programme on Space Applications

The United Nations Programme on Space Applications, established in 1971, has provided support to countries in making full use of the benefits of space technology and its applications for social and economic developments. The Programme has continuously evolved over four decades, taking into account the latest development in the field of space activities, to best serve the capacity-building needs of countries and to help ensure that space-based solutions contribute to improving life on Earth [8].

From 1971 to 2010, the Programme conducted 271 activities such as expert meetings, seminars, workshops, and international conferences in 67 different countries with a total number of 18,251 participants [9]. These activities covered the thematic areas of the Programme: natural resources management, environmental monitoring, enabling space technology, space science, and space law.

III.II Human Space Technology Initiative (HSTI)

HSTI launched by UNOOSA under the framework of the UN Programme on Space Applications builds on the relevant recommendations related to human spaceflight and exploration contained in the report of the UNISPACE III. Those are: (p.70, *"Background and recommendations of the Conference, G. Harnessing the potentials of space at the start of the new millennium*) [10]

*400. International partnerships and cooperation between countries and companies involved in the operation and utilization of the International Space Station and those countries not yet participating in that endeavour should be encouraged.*

*401. Information about utilization of the International Space Station should be disseminated throughout the world in order to increase awareness of the matter in countries not yet participating in that endeavour.*

*402. Mechanisms for improving accessibility from a technical and financial point of view (for example, loans from the World Bank) should be encouraged to simplify utilization of the International Space Station, especially for developing countries."*

The HSTI aims to facilitate access of all interested United Nations Member States to the utilization of the ISS in order to increase the perimeter of countries cooperating on this enterprise, in accordance with its mandates which are:
- To promote international cooperation in human spaceflight and space exploration-related activities;

- To create awareness among Member States on the benefits of utilizing human space technology and its applications;
- To build capacity in microgravity education and research.

III. III HSTI Work Plan

A work plan spanning three years has been developed in order to meet the HSTI objectives. This initiative is based on three sets of actions:
- To provide a forum for exchange of information between ISS partner countries and non-ISS partner countries;
- To inform Member States about microgravity research opportunities on the ISS and other facilities and about on-going and planned research activities;
- To support Member States in increasing their competence level for microgravity research.

The first action comprises the organization of annual expert meetings in different world regions, workshops, and symposiums. This will bring experts together in order to establish a common view on how to facilitate HSTI. These Expert Meetings will be complemented by periodic outreach seminars with the purpose of providing information about HSTI activities to a broader audience. The first one was held in Vienna in February of this year.

The second action involves the publication and distribution of informative materials to United Nations Member States. This will increase awareness of the possibilities and benefits of microgravity science.

The last action includes the distribution of educational materials to researchers and students with the purpose of increasing their capacities in relevant disciplines. The most substantial part of this action will be the distribution of zero-gravity experiment instruments to institutions in developing countries in order to provide them with a low-cost and easy way to conduct research under conditions similar to microgravity.

The principal activities of this work plan are:
(a) To conduct expert meetings and workshops on human space technology;
(b) To develop a zero-gravity experiment instrument distribution project;
(c) To develop educational materials for microgravity science and technology;
(d) To conduct technical advisory services
(e) To conduct scientific outreach activities

Figure 1 shows the timeline of the work plan as currently defined.

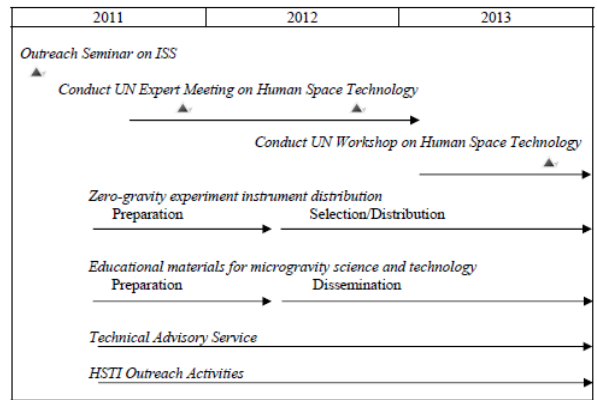

Figure 1: Human Space Technology Initiative Work Plan (2011-2013)

The major milestones are the two United Nations Expert Meetings on Human Space Technology scheduled for the end of this year and of next year as well as the United Nations Workshop on Human Space Technology planned for 2013. Currently, the zero-gravity experiment instrument distribution project and the development of the educational/outreach materials are in their preparatory phase.

IV. HSTI ACTIVITIES

IV. I Outreach Seminar on the ISS on 8 February 2011

As part of the HSTI activities, UNOOSA organized the first Outreach Seminar on the International Space Station (ISS) in Vienna, Austria, on 8 February 2011, in close cooperation with the five ISS partners [11]. For the first time, the United Nations brought together the ISS partners and many non-ISS partner countries, a total of 17, to start discussions on how HSTI could facilitate access to human space research facilities such as the ISS for non-ISS partners. The seminar presented the status of educational and research activities on the ISS and provided information on utilization and cooperation opportunities.

At the seminar, the ISS partners introduced key features of the ISS and its utilization activities including:
- The unique features of the ISS are: a robust, continuous, and sustainable microgravity platform; a continuous human presence in space; access to the ultra-high vacuum of space; a unique altitude for observation and testing; and payload-to-orbit-and-return capability.
- The mechanism of non-ISS partner institutions' participation to ISS utilization, the mechanism of coordination in order to optimize cooperation and avoid duplication in experiments conducted on-board the ISS.

- Respective ISS partners' activities were presented, including, among others, CSA's space robotics, technology development, scientific research, and outreach and education as well as their double approach based on "Push (basic)" knowledge through world-class research on one hand and "Pull (Applied)" innovation on the other; ESA's research assets deployed with the Columbus module for a broad science community and multi-disciplinary/interdisciplinary approach; dedicated collaborative activities of Japan and JAXA using Kibo module with the Asia-Pacific countries through the Asia-Pacific Regional Space Agency Forum (APRSAF) [12]; NASA Authorization Acts of 2005 and 2008 that other U.S. government agencies use the ISS to meet their agencies' objectives and that commercial and non-profit organizations use the ISS in the interest of economic development in space, NASA's ISS Program Office works in three major directions: ISS National Laboratory, NASA-funded Research, and International Partner Integration; Roscosmos carring out a long term research programme and its applications as well as international collaboration projects to build up new relationships and cooperation with developing countries.

In addition, the seminar emphasized the importance of the ISS as an educational platform to develop science and technology.

During the seminar, there were discussions among participants in which they expressed their views including the following:

- What the benefits are that developing countries could gain from the ISS, why and how they want to participate in the ISS and HSTI as well as the importance of creating guidelines for such participation. They also emphasized that the real key would be building local capacity for the advancement of science and technology in support of national priority needs and development goals.
- In order to participate in ISS utilization, a developing country needs to define the research it wants to carry out and the expected outcomes as well as to estimate national readiness to participate in terms of available national professional expertise and funding.
- There would be different ways to involve developing countries in doing research in space and in the utilization of the laboratories on the ISS. The best option would be to identify experiments and research that could be beneficial for meeting national development priorities and current challenges. Capacity building and training of local scientists and researchers would be the prerequisite for this kind of cooperation.
- There is a need to train the younger generation and the importance of developing capacity-building activities in this area. Receiving educational materials from the ISS partners could greatly contribute to boosting that effort, and the United Nations could have a pivotal role in coordinating the distribution of those materials to developing countries.
- The ISS partners work together to plan the full utilization of the ISS. Working groups for this purpose evaluate and coordinate the experiments as they are planned by each partner in order to optimize the plan. There are waiting lists for some agencies and some disciplines needing access to specific facilities. There are opportunities for collaborative research and growth included in these plans, and the ISS partners welcome future dialogue on such collaboration.
- There is a need for UNOOSA to act as a bridge between the ISS partners and developing countries interested in cooperating with the partners by setting up guidelines and specifications for the preparation and design of ISS on-board experiments.

The seminar established that HSTI could be a meaningful mechanism for creating awareness about the potential of the ISS and the research conducted on the ISS among countries, regions, and potential users which have to-date not been involved with such activities, and this would contribute to capacity building in microgravity science and technology education in the world.

IV. II Zero-Gravity Experiment Instrument Demonstration on 9 June 2011

One important building block in the HSTI is the zero-gravity experiment instrument distribution project. Its primary objective is to serve capacity building in developing countries by providing researchers and institutions with a means to conduct microgravity experiments. These instruments may also be used for education in high schools and universities.

There have been quite a few microgravity simulation instruments developed. The most well-known instrument is called a drop tower [13,14], in which an experiment capsule experiences a few seconds of microgravity during a free-fall. Recently, a new instrument called EZ-SPACE [15,16] was developed as a desk-top machine using the similar free-fall concept. These zero-gravity machines can be utilized for carrying out ground-based experiments in fluid physics, combustion, and material science.

As for life science, clinostats [17,18] have been extensively used to observe plant growth and to conduct crystal growth experiments on the ground. A simple clinostat has a rotating horizontal axis, and hence, it can average a direction of the Earth's gravity vector. If a characteristic time of a phenomenon such as plant growth is much longer than the changes of the gravity vector, a clinostat can generate effects similar to that of microgravity. A more complex machine with two rotational axes has been developed in order to produce three-dimensional random rotation [19]. One of these advanced machines is called the Random Positioning Machine [20].

The UNOOSA conducted a demonstration of the desk-top Random Positioning Machine in Vienna, Austria, on 9 June 2011, in cooperation with "Dutch Space," a company from the Netherlands, during the 54th session of COPUOS. Twenty-one delegates from 15 countries participated in the demonstration.

The UNOOSA is currently investigating the most effective way to distribute zero-gravity experiment instruments to institutions in the world. The possible instruments are free-fall type machines for material science experiments and clinostats for life science experiments. The zero-gravity experiment instrument distribution project is expected to start in 2012.

## IV. III Expert Meeting on Human Space Technology on 14-18 November 2011

The first Expert Meeting on Human Space Technology will take place in Malaysia from 14 to 18 November of this year. It will bring ISS partners and experts from United Nations Member States together in order to raise awareness of ISS utilization among Member States and to discuss how to facilitate HSTI activities. Its main topics include research and educational activities in microgravity science and technology, the Non-Partner Participation policy, and of course, the ISS programme itself. It will be organised around sessions for presentations and working groups. During the working group sessions, participants will discuss any issues involving promoting human space technology and how to expand ISS utilization all over the world. The Expert Meeting will conclude with observations and recommendations, which will be reported to the 49th session of the Scientific and Technical Subcommittee of COPUOS to be held in February in 2012.

## V. CONCLUSION

The Human Space Technology Initiative (HSTI), launched by UNOOSA under the framework of the United Nations Programme on Space Applications, aims to promote international cooperation in human spaceflight and space exploration-related activities and to create awareness among United Nations Member States on the benefits of utilizing human space technology and its applications such as the International Space Station.

The major objectives are to disseminate information on microgravity research activities and to increase the level of cooperation between countries in the field of microgravity science and technology and thus further the peaceful uses of outer space for the benefit of all.

In order to increase the capacity to conduct microgravity-related research, a project was set up for the distribution of zero-gravity experiment instruments for research and educational purposes. This project will be accompanied by the distribution of adequately designed educational materials, and thus, allow a considerably larger radius of distribution not available with the limited number of zero-gravity experiment instrument being distributed. The key elements of the HSTI activities will be expert meetings, seminars, and workshops held on a regular basis to serve as a platform for information exchange and encourage cooperation among participating countries and institutions. These meetings, seminars, and workshops will be held in rotating geographical regions in order to achieve the broadest participation in this initiative.

Desirable and expected ways and means for achieving the HSTI objectives will be continuously discussed among participating nations/institutions in the forthcoming HSTI activities.

## ACKNOWLEDGEMENT AND DISCLAIMER

We are very grateful to Mr. Guus Borst of "Dutch Space" from the Netherlands for providing their Random Positioning Machine (RPM) and supporting the demonstration of the RPM during the 54th session of COPUOS in Vienna, Austria, in June of this year.

The views expressed herein are those of the authors and do not necessarily reflect the views of the United Nations.

## REFERENCES[*]

[1] Report of the Committee on The Peaceful Uses of Outer Space, General Assembly Official Records, Twenty-Third Session, New York 1968
[2] General Assembly resolution 2601 A (XXIV). International co-operation in the peaceful uses of outer space; 16 December 1969.
[3] General Assembly resolution 37/90. Second United Nations Conference on the Exploration and Peaceful Uses of Outer Space, para7: redirection of the Programme on Space Applications, regional centres; 10 December 1982.

---

[*] *United Nations documents quoted in this paper are available from the website of the Office for Outer Space Affairs at www.unoosa.org and from the Official Document System of the United Nations at www.documents.un.org*

[4] Report of the Third United Nations Conference on the Exploration and Peaceful Uses of Outer Space, Vienna, 19-30 July 1999, A/CONF.184/6

[5] Reference Guide To The International Space Station, National Aeronautics and Space Administration, November 2010, NASA NP-2010-09-682-HQ

[6] The Era of International Space Station Utilization, NASA, CSA, ESA, JAXA, Roscosmos, NP-2010-03-003-JSC

[7] Jones, R., International Space Station Non-Partner Participation, February 2011, page 81, 82, <http://www.nasa.gov/pdf/542016main_ISS_Overview_HSTI_Feb2011.pdf>; February 2011.

[8] Balogh, W., Canturk, L., Chernikov, S., Doi, T., Gadimova, S., Haubold, H., Kotelnikov, V., The United Nations Programme on Space Applications: Status and direction for 2010, Space Policy 26 (2010) 185e188

[9] Märcz, A., Internship Report on Activities of the United Nations Programme on Space Applications, Office for Outer Space Affairs, 30 June 2011

[10] Report of the Third United Nations Conference on the Exploration and Peaceful Uses of Outer Space, Vienna, 9-30 July 1999, A/CONF.184/6

[11] Human Space Technology Initiative (HSTI) – Report on Outreach Seminar on the International Space Station (ISS) Tuesday, 8 February 2011, Vienna International Centre, A/AC.105/2011/CRP.13, 6 June 2011

[12] Takaoki, M., Fujimoto, N., Ogawa, S., Yamamoto, M., Fujimori, Y., Nagai, N., Miyazaki, K., Kamigaichi, S., Kibo Utilization Cooperation with Asian Countries, ISTS 2011-h-20

[13] Zero Gravity Research Facility User's Guide, NASA/TM-1999-209641, 1999

[14] ZARM Drop Tower User Manual, 2011

[15] Akiba, R., Koreki, T., Egami, I., and Akiba, M., A New Low Gravity Device Applying the Yo-Yo Principle, EZ-SPACE II, JSTS, Vol. 25, No. 1, pp. 1-9.

[16] Koreki, T., Egami, I., and Akiba, R., Educational Experiments with a Low Gravity Device, EZ-SPACE II, JSTS, Vol. 25, No. 1, pp. 10-16.

[17] The Physical Basis of Gravity Stimulus Nullification by Clinostat Rotation, Plant Physiol. (1971) 47, p. 756-764.

[18] Rotation Axes for Clinostat Studies in Light, Plant Physiol. (1070) 45, 231-232.

[19] Hoson, T., Kamisaka, S., Masuda, Y., and Yamashita, M., Changes in Plant Growth Processes under Microgravity Conditions Simulated by a Three-Dimensional Clinostat, Bot. Mag. Tokyo, 105, 1992, pp. 53-70.

[20] Borst, A. G. and van Loon, J. J. W. A., Technology and Development for the Random Positioning Machine, RPM, Microgravity Sci. Technol (2009) 21, pp. 287-292